\documentclass[9pt,shortpaper,twoside,web]{ieeecolor}
\usepackage{mathtools,amsthm}

\usepackage{graphics} 
\usepackage{epsfig}
\usepackage{times} 
\usepackage{amsmath,amssymb,amsfonts}
\usepackage{mathrsfs}
\usepackage{algorithmic}
\let\labelindent\relax

\usepackage{enumitem}
\usepackage{tcolorbox}
\usepackage{graphicx}
\usepackage{textcomp}
\usepackage{varioref}
\usepackage{import}
\usepackage{dsfont}
\usepackage{framed,xcolor}
\usepackage{color}
\usepackage{comment}
\usepackage{multirow}
\usepackage{epstopdf}   
\usepackage{psfrag}
\usepackage{breqn}
\usepackage{float}
\usepackage{mathtools}
\usepackage{booktabs}
\usepackage{soul}
\usepackage{amsbsy}
\usepackage[bottom]{footmisc}
\usepackage{lineno}
\usepackage{cleveref}
\usepackage{microtype}
\usepackage{comment}
\newtheorem{theorem}{Theorem}

\newtheorem{Assumption}{Assumption}
\definecolor{arash}{rgb}{1,0.8,0.8}

\newcommand{\vect}[1]{\ensuremath{\boldsymbol{\mathrm{#1}}}}
\usepackage{graphicx}
\usepackage{tabularx,booktabs}
\usepackage{multicol}
\usepackage{multimedia}
\usepackage{fancybox}
\usepackage{generic}
\usepackage{cite}
\usepackage{amsmath,amssymb,amsfonts}
\usepackage{algorithmic}
\usepackage{graphicx}
\usepackage{textcomp}
\usepackage{soul}
\newcommand{\biggg}{\bBigg@{1.6}}

\definecolor{mario}{rgb}{0.8,0.8,1}

\begin{document}
\bstctlcite{IEEEexample:BSTcontrol}

\title{Equivalence of Optimality Criteria for Markov Decision Process and Model Predictive Control}\author{Arash Bahari Kordabad, Mario Zanon, Sebastien Gros\thanks{Arash Bahari Kordabad and Sebastien Gros are with Department of Engineering Cybernetics, Norwegian University of Science and Technology (NTNU), Trondheim, Norway. Mario Zanon is with the IMT School for Advanced Studies Lucca, Italy. E-mail:{\tt\small arash.b.kordabad@ntnu.no, mario.zanon@imtlucca.it and sebastien.gros@ntnu.no}}}

%\bstctlcite{IEEEexample:BSTcontrol}
\maketitle

%%%%%%%%%%%%%%%%%%%%%%%%%%%%%%%%%%%%%%%%%%%%%%%%%%%%%%%%%%%%%%%%%%%%%%%%%%%%%%%%
\begin{abstract} This paper shows that the optimal policy and value functions of a Markov Decision Process (MDP), either discounted or not, can be captured by a finite-horizon undiscounted Optimal Control Problem (OCP), even if based on an inexact model. This can be achieved by selecting a proper stage cost and terminal cost for the OCP. A very useful particular case of OCP is a Model Predictive Control (MPC) scheme where a deterministic (possibly nonlinear) model is used to reduce the computational complexity. This observation leads us to parameterize an MPC scheme fully, including the cost function. In practice, Reinforcement Learning algorithms can then be used to tune the parameterized MPC scheme. We verify the developed theorems analytically in an LQR case and we investigate some other nonlinear examples in simulations.
\end{abstract}
\IEEEpeerreviewmaketitle

\section{Introduction}
Markov Decision Processes (MDPs) provide a standard framework for the optimal control of discrete-time stochastic processes, where the stage cost and transition probability depend only on the current state and the current input of the system~\cite{puterman2014markov}. A control system, described by an MDP, receives an input at each time instance and proceeds to a new state with a given probability density, and in the meantime, it gets a stage cost at each transition. For an MDP, a policy is a mapping from the state space into the input space and determines how to select the input based on the observation of the current state. This policy can either be a deterministic mapping from the state space \cite{silver2014deterministic} or a conditional probability of the current state, describing the stochastic policy \cite{SuttonPG}. This paper focuses on deterministic policies. Solving an MDP refers to finding an optimal policy that minimizes the expected value of a total cumulative cost as a function of the current state. The cumulative cost can be either discounted or undiscounted with respect to the time instant. Therefore, different definitions for the cumulative cost yields different optimality criteria for the MDPs. Dynamic Programming (DP) techniques can be used to solve MDPs based on the Bellman equations. However, solving the Bellman equations is typically intractable unless the problem is of very low dimension~\cite{bertsekas1995dynamic}. This issue is known as ``curse of dimensionality” in the literature~\cite{powell2007approximate}. Besides, DP requires the exact transition probability of MDPs, while in most engineering applications, we do not have access to the exact probability transition of the real system.

Reinforcement Learning (RL)~\cite{sutton2018reinforcement} and approximate DP~\cite{bertsekas2008approximate} are two common techniques that tackle these difficulties. RL offers powerful tools for tackling MDP without having an accurate knowledge of the probability distribution underlying the state transition. In most cases, RL requires a function approximator to capture the optimal policy or the optimal value functions underlying the MDP. A common choice of function approximator in the RL community is to use a Deep Neural Network (DNN)~\cite{arulkumaran2017deep}. DNNs can be used to capture either the optimal policy underlying the MDP directly or the action-value function from which the optimal policy can be indirectly extracted. However, the formal analysis of closed-loop stability and safety of the policies provided by approximators such as DNNs is challenging. Moreover, DNNs usually need a large number of tunable parameters and a pre-training is often required so that the initial values of the parameters are reasonable.

Model Predictive Control (MPC) is a well-known control strategy that employs a (possibly inaccurate) model of the real system dynamics to produce an input-state sequence over a given finite-horizon such that the resulting predicted state trajectory minimizes a given cost function while explicitly enforcing the input-state constraints imposed on the system trajectories \cite{MPCbook}. For computational reasons, simple models are usually preferred in the MPC scheme. Hence, the MPC model often does not have the structure required to correctly capture the real system dynamics and stochasticity. The idea of using MPC as a function approximator for RL techniques was justified first in~\cite{gros2019data}, where it was shown that the optimal policy of a discounted MDP can be captured by a discounted MPC scheme even if the model is inexact. Recently, MPC has been used in different systems to deliver a structured function approximator for MDPs (see e.g., ~\cite{gros2019data,Arash2021Multi,kordabad2021mpc}) and partially observable MDPs~\cite{Hossein2021MHE}. Stability for discounted MPC schemes is challenging, and for a finite-horizon problem, it is shown in \cite{granzotto2020finite} that even if the provided stage cost, terminal cost and terminal set  satisfy the stability requirements, the closed-loop might be unstable for some discount factors. Indeed, the discount factor has a critical role in the stability of the closed-loop system under the optimal policy of the discounted cost. The conditions for the asymptotic stability for discounted optimal control problems have been recently developed in \cite{zanon2022new} for deterministic systems with the exact model. Therefore, an undiscounted MPC scheme is more desirable, where the closed-loop stability analysis is straightforward and well-developed~\cite{MPCbook}.

The equivalence of MDPs criteria (discounted and undiscounted) has been recently discussed in \cite{zanon2022stability} in the case an exact model of MDP is available. However, in practice, the exact probability transition of the MDP might not be available and  we usually have a (possibly inaccurate) model of the real system. This work extends the results of \cite{zanon2022stability} in the sense of the model mismatch and while extends also the results of~\cite{gros2019data} to the case of using undiscounted MPC scheme to capture a (possibly discounted) MDP. More specifically, we show that, under some conditions, an undiscounted finite-horizon Optimal Control Problem (OCP) can capture the optimal policy and the optimal value functions of a given MDP, either discounted or undiscounted, even if an inexact model is used in the undiscounted OCP. We then propose to use a deterministic (possibly nonlinear) MPC scheme as a particular case of the theorem to formulate the undiscounted OCP as a common MPC scheme. By parameterizing the MPC scheme, and tuning the parameters via RL algorithms one can achieve the best approximation of the optimal policy and the optimal value functions of the original MDP within the adopted MPC structure.

%We will show that if the parameterization of the stage cost and terminal cost are well-posed, then the optimal policy of the MPC-scheme is a stabilizing policy for the model. 

The paper is structured as follows. Section~\ref{sec:MDP} provides the formulation of MDPs under discounted and undiscounted optimality criteria. Section~\ref{sec:MDL} provides formal statements showing that using cost modification in a finite-horizon undiscounted OCP one is able to capture the optimal value function and optimal policy function of the real system with discounted and undiscounted cost even with a wrong model. Section~\ref{sec:MPC} presents a parameterized MPC scheme as a special case of the undiscounted OCP, where the model is deterministic (i.e. the probability transition is a Dirac measure). Then the parameters can be tuned using RL techniques. Section~\ref{sec:analyt} provides an analytical LQR example. Section~\ref{sec:sim} illustrates different numerical simulation.  Finally, section~\ref{sec:cons}  delivers the conclusions.

%finite-horizon Optimal Control Problem (OCP) is able to capture the optimal policy and optimal value functions of the discounted MDP even if an inexact model is used in the OCP.

\section{Real System}\label{sec:MDP}
In this section, we formulate the real system as Markov Decision Processes (MDPs).
We consider an MDP on a continuous state and input spaces over $\mathbb{R}^n$ and $\mathbb{R}^m$, respectively, with stochastic states $\vect{s}_k\in\mathcal{X}\subseteq\mathbb{R}^{n}$ in the Lebesgue-measurable set $\mathcal{X}$ and inputs $a_k\in \mathcal{U}\in \mathbb{R}^m$. The triple $(\Omega, \mathcal{F}, \rho)$ defines the probability space associated with a Markov chain, where $\Omega=\Pi_{k=0}^\infty \mathcal{X}$, with associated $\sigma$-field $\mathcal{F}$ and $\rho$ is the probability measure. We then consider stochastic dynamics defined by the following conditional
probability measure:
\begin{align}\label{eq:MC}
\rho\left[{\vect s}_{k+1}|\vect s_k,\vect a_k\right]\,,    
\end{align}
defining the conditional probability of observing a transition from a given state-action pair $\vect{s}_k$, $\vect{a}_k$ to a subsequent state $\vect{s}_{k+1}$. The input $\vect a$ applied to the system for a given state $\vect s$ is selected by a deterministic policy $\vect \pi : \mathcal{X} \rightarrow \mathcal{U}$. We denote ${\vect s}^{\vect\pi}_{0,1,\ldots}$ the (possibly stochastic) trajectories of the system \eqref{eq:MC} under policy $\vect\pi$, i.e., ${\vect s}^{\vect\pi}_{k+1}\sim \rho\left[\cdot|{\vect s}^{\vect\pi}_{k},\vect \pi({\vect s}^{\vect\pi}_{k})\right]$, starting from ${\vect s}^{\vect\pi}_{0}=\vect s$, $\forall \vect \pi$. We further denote the measure associated with such trajectories as $\tau_k^{\vect{\pi}}$ in the same space as $\rho$. More specifically, $\tau_0^{\vect{\pi}}(\cdot)=\rho_0(\cdot),\,\forall \vect\pi$,  where $\rho_0(\cdot)$ is the initial state
distribution and $ 
    \tau_{k+1}^{\vect{\pi}}(\cdot):=\int_{\mathcal{X}}\rho\left[\cdot|\vect s,\vect \pi(\vect s)\right]\tau_{k}^{\vect{\pi}}(\mathrm{d}\vect s)\,,k>0.$
\subsection{Discounted MDPs}

In the discounted setting, we aim to find the optimal policy $\vect\pi^\star$, solution of the following discounted infinite-horizon OCP:% the following problem: 
\begin{align}\label{eq:value1}
V^\star(\vect s):=\min_{\vect\pi}&\, V^{\vect\pi}(\vect s):=\mathbb{E}_{\tau^{\vect{\pi}} } \left[ \sum_{k=0}^\infty\, \gamma^k \ell(\vect s^{\vect\pi}_k,\vect { \pi}\left(\vect s^{\vect\pi}_k\right))\right],
\end{align}
for all initial states $\vect s^{\vect\pi}_0=\vect s$, where $V^\star:\mathcal{X}\rightarrow\mathbb{R}$ is the optimal value function, $V^{\vect\pi}$ is the value function of the Markov Chain in closed-loop with policy $\vect\pi$, $\ell:\mathcal{X}\times\mathcal{U}\rightarrow \mathbb{R}$ is the stage cost function of the real system and $\gamma\in(0,1]$ is the discount factor. The expectation $\mathbb{E}_{\tau^{\vect{\pi}} }$ is taken over the distribution underlying the Markov Chain \eqref{eq:MC} in closed-loop with policy $\vect\pi$, i.e., $\vect{s}_k\sim \tau_k^{\vect{\pi}}(\cdot)$ for $k>0$. The action-value function $Q^\star(\vect s,\vect a)$ and advantage function $A^\star(\vect s,\vect a)$ associated to \eqref{eq:value1} are defined as follows:
\begin{subequations}
\begin{align}
    Q^\star(\vect s,\vect a)&:=\ell(\vect s,\vect a)+\gamma \mathbb{E}_\rho\left[V^\star({\vect s}^+)|\vect s,\vect a\right], \\
    A^\star(\vect s,\vect a)&:= Q^\star(\vect s,\vect a)-V^\star(\vect s).
\end{align}
\end{subequations}
Then from the Bellman equation, we have the following identities:
\begin{subequations}
\begin{align}
&V^\star(\vect s)= Q^\star(\vect s,\vect \pi^\star(\vect s))=\min_{\vect a} Q^\star(\vect s,\vect a) , \quad \forall \vect s\in \mathcal{X},\label{eq:Bell}\\
\label{eq:Bell:A}
    0&=\min_{\vect a} A^\star(\vect s,\vect a),\,\,\,\,
    \vect\pi^\star(\vect s)\in \mathrm{arg}\min_{\vect a} A^\star(\vect s,\vect a),\,\, \forall \vect s\in \mathcal{X}.
\end{align}
\end{subequations}

\subsection{Undiscounted MDPs}\label{sec:undis}
Undiscounted MDPs refer to MDPs when $\gamma=1$. In this case $V^\star$ is in general unbounded and the MDP is ill-posed. In order to tackle this issue, alternative optimality criteria are needed. Gain optimality is one of the common criteria in the undiscounted setting. Gain optimality is defined based on the following average-cost problem:
\begin{align}\label{eq:gain}
\bar V^\star(\vect s):=\min_{\vect\pi} \lim_{N\rightarrow\infty}\frac{1}{N}\mathbb{E}_{\tau^{\vect{\pi}} }\left[ \sum_{k=0}^{N-1}\, \ell(\vect s^{\vect\pi}_k,\vect { \pi}\left(\vect s^{\vect\pi}_k\right))\right],
\end{align}
for all initial states $\vect s^{\vect\pi}_0=\vect s$, $\forall\vect\pi$, where $\bar V^\star$ is the optimal average cost. We denote the optimal policy solution of \eqref{eq:gain} as $\bar {\vect\pi}^\star$. This optimal policy is called  \textit{gain optimal}. The gain optimal policy $\bar {\vect\pi}^\star$ may not be unique. Moreover, the optimal average cost  $\bar V^\star$ is commonly  assumed to be independent of the initial state $\vect s$~\cite{mahadevan1996average}. This assumption e.g. holds for \textit{unichain} MDPs, in which under any policy any state can be reached in finite time from any other state. Unfortunately, the gain optimality criterion only considers the optimal steady-state distribution and it overlooks transients. As an alternative, \textit{bias optimality} considers the optimality of the transients. Precisely, bias optimality can be formulated through the following OCP:
\begin{align}\label{eq:value6}
\tilde V^\star(\vect s)=\min_{\vect\pi} \mathbb{E}_{\tau^{\vect{\pi}} }\left[ \sum_{k=0}^\infty\, (\ell(\vect s^{\vect\pi}_k,\vect { \pi}\left(\vect s^{\vect\pi}_k\right))-\bar V^\star)\right],
\end{align}
%   We denote the bias optimal policy solution of \eqref{eq:value6} by $\tilde {\vect\pi}^\star$.
where $\tilde V^\star$ is the optimal value function associated to bias optimality. Note that \eqref{eq:value6} can be seen as a special case of the discounted setting in \eqref{eq:value1} when $\gamma=1$ and the optimal average cost $\bar V^\star$ is subtracted from the stage cost in \eqref{eq:value1}. Therefore, for the rest of the paper we will consider the discounted setting \eqref{eq:value1}. Without loss of generality we assume that  $\bar V^\star=0$ in the case $\gamma=1$. This choice yields a well-posed optimal value function in the undiscounted setting. Clearly, if this does not hold, one can shift the stage cost to achieve $\bar V^\star=0$.

%We then define the corresponding action-value function and advantage function as follows:
%\begin{subequations}\begin{align}    \tilde Q^\star(\vect s,\vect a):&=\ell(\vect s,\vect a)-\bar V^\star+ \mathbb{E}_\rho\left[\tilde V^\star({\vect s}^+)|\vect s,\vect a\right]\\    \tilde A^\star(\vect s,\vect a):&= \tilde Q^\star(\vect s,\vect a)-\tilde V^\star(\vect s)
%\end{align}\end{subequations}
%Then the Bellman equations reads:
%\begin{subequations}\begin{align}    0&=\min_{\vect \pi} \tilde A^\star(\vect s,\vect \pi(\vect s)), \\ \tilde {\vect\pi}^\star&=\mathrm{arg}\,\min_{\vect \pi} \tilde A^\star(\vect s,\vect \pi(\vect s))
%\end{align}\end{subequations}
%\begin{Corollary} Using Theorem \ref{theorem1}, Under assumption \ref{assume2} for the optimal value function $\tilde V^\star$, One can show that there exists a stage cost $\hat L:\mathcal{X}\times\mathcal{U}\rightarrow \mathbb{R}$ such that the undiscounted infinite-horizon OCP with model \eqref{eq:model} in \eqref{eq:value2} delivers the bias optimal policy and value functions, $\forall \vect s\in\mathcal{S}$ with the real system .\end{Corollary}

%In this section it is shown that undiscounted finite-horizon OCP can capture the optimal value of an MDP with gain and bias criteria even if an inaccurate model is used by selecting a proper stage cost and terminal cost. 
\section{Model of the system}\label{sec:MDL}
In general, we may not have full knowledge of the probability transition of the real MDP \eqref{eq:MC}. One then typically considers an imperfect model of the real MDP \eqref{eq:MC}, having the state transition:
\begin{align}\label{eq:model}
\hat\rho\left[{\vect s}_{k+1}|\vect s_k,\vect a_k\right].
\end{align}
in the same space as $\rho$. In order to distinguish it from the real system trajectory, let us denote $\hat{\vect s}^{\vect\pi}_{0,1,\ldots}$ the (possibly stochastic) trajectories of the state transition model \eqref{eq:model} under policy $\vect\pi$, i.e., $\hat{\vect s}^{\vect\pi}_{k+1}\sim \hat\rho\left[\cdot|\hat{\vect s}^{\vect\pi}_{k},\vect \pi(\hat{\vect s}^{\vect\pi}_{k})\right]$, starting from $\hat{\vect s}^{\vect\pi}_{0}=\vect s$, $\forall \vect\pi$. We further denote the measure associated with such trajectories as $\hat \tau^{\vect{\pi}}$. In general, $\tilde{\cdot}$ refers to the notations related to the imperfect model of the system in this paper. It has been shown in \cite{gros2021dissipativity} that proving closed-loop stability of the Markov Chains with the optimal policy resulting from an undiscounted OCP is more straightforward than a discounted setting~\cite{zanon2022stability}. This observation is well-known in MPC of deterministic systems \cite{postoyan2016stability}. Therefore, in this paper, we are interested in using an undiscounted OCP for the model \eqref{eq:model} in order to extract the optimal policy and optimal value functions of the real system \eqref{eq:MC}, as this allows us to enforce stability guarantees. 

\subsection{Finite-horizon OCP}
While MPC allows one to introduce stability and safety guarantees, it also requires a model of the real system which is bound to be imperfect, and it optimizes the cost over a finite horizon with unitary discount factor. In other words, MPC is an MDP based on the imperfect system model~\eqref{eq:model} which we will formulate in~\eqref{eq:value2}. In this section we will prove that these differences between the MPC formulation and the original MDP formulation do not hinder the ability to obtain the optimal policy and the optimal value functions of the real system through MPC. Consider the following undiscounted finite-horizon OCP associated to model \eqref{eq:model}:
\begin{align}\label{eq:value2}
\hat V_N^\star(\vect s)=\min_{\vect\pi}\,\hat V_N^{\vect\pi}(\vect s):=&\mathbb{E}_{\hat \tau^{\vect{\pi}}}\bigg[  \hat T(\hat{\vect s}^{\vect\pi}_N)+\sum_{k=0}^{N-1}\, \hat L(\hat {\vect s}^{\vect\pi}_k,\vect { \pi}\left(\hat {\vect s}^{\vect\pi}_k\right))\bigg],
\end{align}
with initial state $\hat {\vect s}^{\vect\pi}_0=\vect s$, where $N\in\mathbb{N}$ is the horizon length, $ \hat T$, $\hat L$, $\hat V_N^\star$ and $\hat V_N^{\vect\pi}$ are the terminal cost, the stage cost, the optimal value function and the value function of the policy $\vect\pi$ associated to model \eqref{eq:model}, respectively, and where $\mathbb{N}$ is the set of natural numbers. The expectation $\mathbb{E}_{\hat \tau^{\vect{\pi}}}$ in \eqref{eq:value2} is taken over undiscounted closed-loop Markov Chain \eqref{eq:model} with policy $\vect\pi$. We denote $\hat{\vect\pi}_N^\star$ the optimal policy resulting from \eqref{eq:value2}.
Moreover, the action-value function $ \hat Q_N^\star$ associated to \eqref{eq:value2} is defined as follows:
\begin{subequations}
\begin{align}\label{eq:hatQ}
    \hat Q_N^\star(\vect s,\vect a)&:=\hat L(\vect s,\vect a)+\mathbb{E}_{\hat\rho}\left[ \hat V_{N-1}^\star({\vect s}^+)|\vect s,\vect a\right]\,,\\
    \hat V_{0}^\star({\vect s})&:=\hat T({\vect s})
\end{align}
\end{subequations}

\par The next assumption expresses a requirement on the boundedness of $V^\star$ under model trajectories $\hat{\vect s}^{\vect\pi}_{0,1,\ldots}$ with the optimal policy $\vect\pi^\star$ which allows us to develop the theoretical results of this paper.

\begin{Assumption}\label{assume:S} The following set is non-empty for a given $\bar N\in\mathbb{N}$.%following sets $\mathcal{S}$:
\begin{align}\label{eq:assum1}
\mathcal{S}&=:\left\{\vect s\in\mathcal{X}\,\,\Big|\,\,\left|\mathbb{E}_{\hat \tau^{\vect{\pi}^{\star}}} \left[V^\star(\hat{\vect s}^{\vect\pi^\star}_k)\right]\right|<\infty, \ \forall\, k \leq \bar N\right\}
\end{align}
\end{Assumption}
Assumption~\ref{assume:S} requires that there exists a non-empty set $\mathcal{S}$ such that for all trajectories starting in it, the expected value of $V^\star$ is bounded at all future times under the state distribution given by the model in finite time under the optimal policy. 
%for which if the initial state $\hat{\vect s}^{\vect\pi}_{0}=\vect s$ starts form this set, the expected value of the optimal value function for the real system, defined in \eqref{eq:value1}, remains finite in a finite-horizon under the model trajectories $\hat{\vect s}^{\vect\pi^\star}_k$. 
This assumption plays a vital role in the derivation of our main result. We will further detail this assumption in Section~\ref{sec:assum}.

The next theorem provides theoretical support to the idea that one can recover the optimal policy and value functions by means of an MPC scheme which is based on an imperfect model and has an undiscounted formulation over a finite prediction horizon.
%states that, under assumption \ref{assume:S}, an undiscounted finite-horizon OCP yields the same optimal value function $V^\star$, optimal action-value function $Q^\star$ and optimal policy $\vect\pi^\star$ for the real MDP $\rho$ with the discounted cost even if using an inaccurate model $\hat\rho$. This idea first was established in \cite{gros2019data} when the model uses discounted finite-horizon OCP, and we generalize it here to the undiscounted setting, and we show that an undiscounted setting is able to capture the discounted optimal value functions independent of the discount factor $\gamma$.

\begin{theorem}\label{theorem1} Suppose that Assumption~\ref{assume:S} holds for $\bar N \geq N$. Then, there exist a terminal cost $\hat T$ and a stage cost $\hat L$ such that the following identities hold, $\forall\, \gamma$, $N\in\mathbb{N}$ and $\vect s\in\mathcal{S}$:
\begin{enumerate}[label=(\roman*)]
  \item $\hat{\vect\pi}_N^\star(\vect s)=\vect\pi^\star(\vect s), $ \label{eq:pipi}
  \item $\hat V_N^\star(\vect s)=V^\star(\vect s),\,\, $\label{eq:VV}
  \item $\hat Q_N^\star(\vect s,\vect a)=Q^\star(\vect s,\vect a),\,\,$ for the inputs $\vect a\in\mathcal{U}$ such that $\lvert\mathbb{E}_{\hat\rho}\left[V^\star({\vect s}^{+})|\vect s,\vect a\right]\rvert<\infty$\label{eq:QQ}
\end{enumerate}
\end{theorem}
\begin{proof}
We select the terminal cost $\hat T$ and the stage cost $\hat L$ as follows:%
\begin{subequations}\label{eq:TL}
\begin{align}
&\null\qquad\qquad\qquad\quad\qquad\hat T(\vect s)= V^\star (\vect s) \label{eq:That0}\\
  &\hat L(\vect s,\vect a)=\label{eq:lhat0}\\ &\left\{\begin{matrix}
 Q^\star(\vect s,\vect a)- \mathbb{E}_{\hat\rho}\left[V^\star({\vect s}^+)|\vect s,\vect a\right] & \mathrm{If}\, \left|\mathbb{E}_{\hat\rho}\left[V^\star({\vect s}^+)|\vect s,\vect a\right]\right|<\infty \\ \infty
 & \mathrm{otherwise}
\end{matrix}\right.\nonumber
\end{align}
\end{subequations}
Under Assumption~\ref{assume:S}, the terminal and stage costs in \eqref{eq:value2} have a finite expected value for all $\vect{\hat s}_0^{\vect{\pi}^{\star}}\in\mathcal{S}$. By substitution of \eqref{eq:TL} in \eqref{eq:value2} and using telescopic sum, we have:
\begin{align}\label{eq:hat:V:pi}
&\hat V_N^{\vect\pi}(\vect s)\nonumber\\
&\hspace{0.5em}=\mathbb{E}_{\hat \tau^{\vect{\pi}}}\bigg[  \hat T(\hat{\vect s}^{\vect\pi}_N)+
\sum_{k=0}^{N-1}\, \hat L(\hat {\vect s}^{\vect\pi}_k,\vect { \pi}\left(\hat {\vect s}^{\vect\pi}_k\right))\bigg]\nonumber\\
&\hspace{0.1em}\overset{\eqref{eq:TL}}{=}\mathbb{E}_{\hat \tau^{\vect{\pi}}}\bigg[  V^\star(\hat{\vect s}^{\vect\pi}_N) +
\sum_{k=0}^{N-1}\, \Big( Q^\star(\hat {\vect s}^{\vect\pi}_k,\vect { \pi}\left(\hat {\vect s}^{\vect\pi}_k\right))-
V^\star(\hat {\vect s}^{\vect\pi}_{k+1})\Big)\bigg]\nonumber\\
&\hspace{0.5em}=Q^\star(\vect s,\vect\pi(\vect s))+\mathbb{E}_{\hat \tau^{\vect{\pi}}}\left[  \sum_{k=1}^{N-1}\, \left( Q^\star(\hat {\vect s}^{\vect{\pi}}_k,\vect { \pi}\left(\hat {\vect s}^{\vect \pi}_k\right))-V^\star(\hat {\vect s}^{\vect \pi}_k) \right)\right]\nonumber\\
&\hspace{0.5em}=Q^\star(\vect s,\vect\pi(\vect s))+\mathbb{E}_{\hat \tau^{\vect{\pi}}}\left[  \sum_{k=1}^{N-1}\, A^\star(\hat {\vect s}^{\vect \pi}_k,\vect { \pi}\left(\hat {\vect s}^{\vect \pi}_k\right))\right],
\end{align}
where $\hat {\vect s}_0=\vect s$. From \eqref{eq:Bell} and \eqref{eq:Bell:A}, we know that:
\begin{align}
    \vect\pi^\star(\cdot)=\arg\min_{\vect{\pi}}A^\star\left(\cdot,\vect { \pi}\left(\cdot\right)\right)=\arg\min_{\vect{\pi}}Q^\star\left(\cdot,\vect { \pi}\left(\cdot\right) \right)
\end{align}
then from \eqref{eq:hat:V:pi}:
\begin{align}
    \vect\pi^\star(\vect s)&=\arg\min_{\vect{\pi}}\hat V_N^{\vect\pi}(\vect s) \\
    &=\arg\min_{\vect{\pi}}Q^\star(\vect s,\vect\pi(\vect s))+\mathbb{E}_{\hat \tau^{\vect{\pi}}}\left[  \sum_{k=1}^{N-1}\, A^\star(\hat {\vect s}^{\vect \pi}_k,\vect { \pi}\left(\hat {\vect s}^{\vect \pi}_k\right))\right]\nonumber
\end{align}
    Note that $\vect\pi^\star$ minimizes all terms in the cost above, i.e., $A^\star$ and $Q^\star$, such that is must also minimize $\hat V_N^{\vect\pi}$. This proves \ref{eq:pipi}, i.e.,
\begin{align*}
    \vect\pi^\star(\vect s)=\hat{\vect\pi}_N^\star(\vect s).
\end{align*}
In turn, this proves \ref{eq:VV}, since
\begin{align}
\hat V^{\star}_N(\vect s)&=\hat V_N^{\vect\pi^\star}(\vect s)=Q^\star(\vect s,\vect\pi^\star(\vect s))+\nonumber\\&+\mathbb{E}_{\hat \tau^{\vect\pi}}\left[  \sum_{k=1}^{N}\, \underbrace{A^\star(\hat {\vect s}^{\vect{\pi}^\star}_k,\vect { \pi}^\star\left(\hat {\vect s}^{\vect{\pi}^\star}_k\right))}_{\stackrel{(\ref{eq:Bell:A})}=0}\bigg| \hat {\vect s}_0=\vect s\right]\nonumber\\&{=}Q^\star(\vect s,\vect\pi^\star(\vect s))\stackrel{(\ref{eq:Bell})}=V^\star(\vect s).
\end{align}
Moreover, from \eqref{eq:hatQ} and \eqref{eq:lhat0}, for any inputs $\vect a\in\mathcal{U}$ such that $\lvert\mathbb{E}_{\hat\rho}\left[V^\star({\vect s}^{+})|\vect s,\vect a\right]\rvert<\infty$, we have:
\begin{align}\label{eq:Qhat:Q}
    &\hat Q_N^\star(\vect s,\vect a)=\hat L(\vect s,\vect a)+\mathbb{E}_{\hat\rho}\left[ \hat V_{N-1}^\star({\vect s}^+)|\vect s,\vect a\right]\\&\stackrel{(\ref{eq:lhat0})}=Q^\star(\vect s,\vect a)+\mathbb{E}_{\hat\rho}\left[ \hat V_{N-1}^\star({\vect s}^+)- V^\star({\vect s}^+)|\vect s,\vect a\right]=Q^\star(\vect s,\vect a),\nonumber
\end{align}
where the last inequality is obtained by noting that \ref{eq:VV} for $N>1$ and $\hat V_{0}^\star({\vect s})=\hat T({\vect s})=V^\star({\vect s})$ for $N=1$. This directly yields \ref{eq:QQ}.
\end{proof}
Theorem~\ref{theorem1} states that, independent of the discount factor $\gamma$, it is possible to find a finite-horizon OCP cost function that provides the optimal policy and optimal value functions of a discounted MDP if an inexact model is used in the finite-horizon OCP. 
We observe that the setup of this paper has been analyzed in~\cite{zanon2022stability}, under the assumption of a perfect model, i.e., $\hat \rho[\cdot|\vect s,\vect a]=\rho[\cdot|\vect s,\vect a]$. In that case \eqref{eq:lhat0} reads:
\begin{align}\label{eq:cost:mod}
   \hat{L}(\vect s,\vect a)=\ell(\vect s,\vect a)+(\gamma-1)\mathbb{E}_{\rho}[V^\star(\vect s^+)|\vect s,\vect a],\, \forall \vect s\in \mathcal{S}, 
\end{align}
which corresponds to the cost modification discussed in~\cite{zanon2022stability}.

%Moreover, if the model $\hat\rho$ is exact, i.e., if $\hat \rho[\cdot|\vect s,\vect a]=\rho[\cdot|\vect s,\vect a]$, then \eqref{eq:lhat0} reads:
%\begin{align}\label{eq:cost:mod}
%   \hat{L}(\vect s,\vect a)=\ell(\vect s,\vect a)+(\gamma-1)\mathbb{E}_{\rho}[V^\star(\vect s^+)|\vect s,\vect a],\, \forall \vect s\in \mathcal{S} 
%\end{align}
%We observe that this is the result obtained in~\cite{zanon2021stability}, such that our result is a generalization.
%Note that our result is a generalization of~\cite{zanon2021stability}, where the case of a perfect model, i.e., $\hat \rho[\cdot|\vect s,\vect a]=\rho[\cdot|\vect s,\vect a]$ and \eqref{eq:cost:mod}, is analyzed.
%\mario{this is a repetition of what we just wrote}
%Note that our result is an extension of~\cite{zanon2021stability}, where it was shown that the stage cost modification in \eqref{eq:cost:mod} yields a discounted MDP which is equivalent to an undiscounted MDP in case $\hat \rho[\cdot|\vect s,\vect a]=\rho[\cdot|\vect s,\vect a]$, i.e., the model is exact, while our result also holds for an inexact model.
\subsection{Infinite-horizon OCP}
In this section, we investigate the case $N\rightarrow \infty$ for which, under
some conditions, the terminal cost can be dismissed. In this case, we first make the next additional assumption.
\begin{Assumption}\label{assume2}
We assume that the optimal value function converges to a constant and finite value with model \eqref{eq:model} under the optimal policy $\vect\pi^\star$. I.e.:
\begin{align}\label{eq:shat:inf}
    -\infty<\lim_{N\rightarrow\infty} \mathbb{E}_{\hat \tau^{\vect{\pi}^{\star}}}\left[  V^\star(\hat{\vect s}^{\vect\pi^\star}_{N})\right]=\hat v_{\infty}<\infty
\end{align}
\end{Assumption}
Assumption~\ref{assume2} can be interpreted as some forms of the stability condition on the model dynamics under the optimal policy $\vect\pi^\star$. We will explain this assumption in Section~\ref{sec:assum}.
In this section, we consider the following undiscounted value function without terminal cost:
\begin{align}\label{eq:inf:OCP} 
\hat V_{\infty}^\star(\vect s):=\min_{\vect\pi}\hat V_{\infty}^{\vect\pi}(\vect s):=\lim_{N\rightarrow \infty}\mathbb{E}_{\hat\tau^{\vect\pi}}\bigg[ \sum_{k=0}^{N-1} \hat L(\hat {\vect s}^{\vect\pi}_k,\vect { \pi}\left(\hat {\vect s}^{\vect\pi}_k\right))\bigg]
\end{align} 
with initial state $\hat {\vect s}^{\vect\pi}_0=\vect s$. We denote the optimal policy solution of \eqref{eq:inf:OCP} as $\hat{\vect\pi}_\infty^\star(\vect s)$. We then define the optimal action-value function $\hat Q_{\infty}^\star$ associated to \eqref{eq:inf:OCP} as follows:
\begin{align}
    \hat Q_{\infty}^\star(\vect s,\vect a)= \hat L(\vect s,\vect a)+\mathbb{E}_{\hat\rho}\left[\hat V_{\infty}^\star(\vect s^+)|\vect s,\vect a\right]\,\,,
\end{align}
We are now ready to state the equivalent of Theorem~\ref{theorem1} in case of an infinite horizon without a terminal cost. 
\begin{theorem}\label{theroem2}  
Suppose that Assumptions~\ref{assume:S} and \ref{assume2} hold, then the following hold $\forall \vect s\in\mathcal{S},\forall \gamma$:
\begin{enumerate}[label=(\roman*)]
  \item $\hat{\vect\pi}_\infty^\star(\vect s)=\vect\pi^\star(\vect s)$\label{eq:pipi2}
  \item $\hat V_{\infty}^\star(\vect s)=V^\star(\vect s)-\hat v_{\infty}$\label{eq:vv2}
  \item $\hat Q_{\infty}^\star(\vect s,\vect a)=Q^\star(\vect s,\vect a)-\hat v_{\infty}$, for the inputs $\vect a\in\mathcal{U}$ such that $\lvert\mathbb{E}_{\hat\rho}\left[V^\star({\vect s}^{+})|\vect s,\vect a\right]\rvert<\infty$\label{eq:QQ2}
\end{enumerate}
if the stage cost $\hat{L}$ is selected according Equation~\eqref{eq:lhat0}.
\end{theorem}
\begin{proof} Using stage cost $\hat{L}$ in \eqref{eq:lhat0}, we have:
\begin{align}
\hat V_{\infty}^{\vect\pi}(\vect s)&=\lim_{N\rightarrow \infty}\,\,\mathbb{E}_{\hat\tau^{\vect\pi}}\bigg[ \sum_{k=0}^{N-1}\, Q^\star(\hat {\vect s}^{\vect\pi}_k,\vect { \pi}\left(\hat {\vect s}^{\vect\pi}_k\right))-\\&\qquad\qquad\qquad\mathbb{E}_{\hat\rho}\left[V^\star(\hat {\vect s}^{\vect\pi}_{k+1})|\hat {\vect s}^{\vect\pi}_k,\vect { \pi}\left(\hat {\vect s}^{\vect\pi}_k\right)\right]\bigg]\nonumber\\
&=\lim_{N\rightarrow \infty}\,\,\mathbb{E}_{\hat\tau^{\vect\pi}}\bigg[ \sum_{k=0}^{N-1}\, Q^\star(\hat {\vect s}^{\vect\pi}_k,\vect { \pi}\left(\hat {\vect s}^{\vect\pi}_k\right))-V^\star(\hat {\vect s}^{\vect\pi}_{k+1})\bigg]\nonumber\\
&=Q^\star(\vect s,\vect \pi(\vect s))+\lim_{N\rightarrow \infty} \mathbb{E}_{\hat\tau^{\vect\pi}}\Bigg[-V^\star(\hat {\vect s}^{\vect\pi}_N)+\nonumber\\&\qquad\qquad\sum_{k=1}^{N-1} Q^\star(\hat {\vect s}^{\vect\pi}_k,\vect\pi(\hat {\vect s}^{\vect\pi}_k))- V^\star(\hat {\vect s}^{\vect\pi}_k)\Bigg]\nonumber\\
&\hspace{-3em}=Q^\star(\vect s,\vect \pi(\vect s))+\lim_{N\rightarrow \infty} \mathbb{E}_{\hat\tau^{\vect\pi}}\Bigg[-V^\star(\hat {\vect s}^{\vect\pi}_N)+\sum_{k=1}^{N-1} A^\star(\hat {\vect s}^{\vect\pi}_k,\vect\pi(\hat {\vect s}^{\vect\pi}_k))\Bigg]\nonumber
\end{align}
where $\hat {\vect s}^{\vect\pi}_0=\vect s$. % and we have used telescopic sum. 
By \eqref{eq:Bell} and \eqref{eq:Bell:A} we know that the policy $\vect\pi(\vect s)=\vect\pi^\star(\vect s)$ minimizes  all terms $A^\star(\cdot,\vect\pi(\cdot))$ and $Q^\star(\cdot,\vect\pi(\cdot))$, such that it also minimizes $\hat V_{\infty}^{\vect\pi}(\vect s)$, i.e.,:
\begin{align}
    \hat{\vect\pi}_\infty^\star(\vect s)=\mathrm{arg}\min_{\vect\pi}\hat V_{\infty}^{\vect\pi}(\vect s)=\vect\pi^\star(\vect s)\,,
\end{align}
which proves \ref{eq:pipi2}. Moreover:
\begin{align}
    \hat V_{\infty}^{\vect\pi^\star}(\vect s)=V^\star(\vect s) -\lim_{N\rightarrow \infty} \mathbb{E}\left[V^\star(\hat {\vect s}^{\vect\pi^\star}_N)\right].
\end{align}
Using \eqref{eq:shat:inf} we have:
\begin{align}
    \hat V_{\infty}^\star(\vect s)=\hat V_{\infty}^{\vect\pi^\star}(\vect s)=V^\star(\vect s)-\hat v_\infty.
\end{align} 
For the inputs $\vect a\in\mathcal{U}$ such that $\lvert\mathbb{E}_{\hat\rho}\left[V^\star({\vect s}^{+})|\vect s,\vect a\right]\rvert<\infty$:
\begin{align}
    &\hat Q_{\infty}^\star(\vect s,\vect a)= \hat L(\vect s,\vect a)+\mathbb{E}_{\hat\rho}\left[\hat V_{\infty}^\star(\vect s^+)|\vect s,\vect a\right]\\
    &=Q^\star(\vect s,\vect a)-\mathbb{E}_{\hat\rho}\left[ V^\star(\vect s^+)|\vect s,\vect a\right]+\mathbb{E}_{\hat\rho}\left[\hat V_{\infty}^\star(\vect s^+)|\vect s,\vect a\right]\nonumber\\
    &=Q^\star(\vect s,\vect a)-\mathbb{E}_{\hat\rho}\left[ V_{\infty}^\star(\vect s^+)-\hat V^\star(\vect s^+)|\vect s,\vect a\right]=Q^\star(\vect s,\vect a)-\hat v_{\infty},\nonumber
\end{align}
which completes the proof.
\end{proof}
Theorem~\ref{theroem2} extends Theorem~\ref{theorem1} to the case of an infinite horizon with zero terminal cost. Assumption~\ref{assume2} is necessary in order to be able to remove the terminal cost. In the next section we will detail the use of the theorems in practice and reformulate OCP \eqref{eq:value2} as a Model Predictive Control (MPC) scheme.
\section{MPC as a function approximator for RL}\label{sec:MPC}
As it was shown in the previous section, the optimal policy and value functions of any MDP with either discounted or undiscounted criteria can be captured using a finite-horizon undiscounted OCP \eqref{eq:value2} even if the model is not accurate. Since the equivalence only holds at the initial state, if one is interested in recovering the optimal MDP policy, the finite-horizon OCP needs to be solved from scratch for each initial state. In practice, this amounts to deploying the finite-horizon OCP in an MPC framework, i.e., in a closed-loop.

As discussed above, the equivalence is only obtained if a properly modified stage and terminal costs are introduced for the finite-horizon undiscounted MPC scheme. However, finding such costs requires knowledge about the optimal value functions of the real MDP. In this section, we detail how the theorems we provided in the previous sections can be used in practice to exploit MPC as a structured function approximator of the optimal policy and value functions of the real MDP. One of the main advantages of MPC is that it allows us to straightforwardly introduce state and input constraints in the policy. We parameterize the MPC scheme with parameter vector~$\vect\theta$ such that RL methods can be deployed to tune $\vect\theta$ in order to achieve the equivalence yielding the optimal policy and value functions of the real system and, consequently, the best possible closed-loop performance.

As the MPC model is not required to  capture the real system dynamics exactly, for the sake of reducing the computational burden, and due to the (relative) simplicity of the resulting MPC scheme, a popular choice of model $\hat\rho\left[{\vect s}^+|\vect s,\vect a\right]$ is a deterministic model, i.e.:
\begin{align}\label{eq:det:model}
    \hat\rho\left[{\vect s}^+|\vect s,\vect a\right]=\delta\left({\vect s}^+-\vect f_{\vect\theta}(\vect s,\vect a)\right)
\end{align}
where $\delta(\cdot)$ is the Dirac measure and $\vect f_{\vect\theta}(\vect s,\vect a)$ is a parameterized deterministic (possibly nonlinear) model. We approximate the modified costs $\hat L$ and $\hat T$ by parametric functions $L_{\vect\theta}$ and $T_{\vect\theta}$, respectively. 
%We then consider the following ENMPC-scheme:
%\begin{subequations}\label{eq:MPC}
%\begin{align}
 %   \min_{\hat{\vect a}}\,\,& T_{\vect\theta}(\hat {\vect s}_N)+\sum_{k=0}^{N-1}L_{\vect\theta} (\hat{ \vect s}_k,\hat{ \vect a}_k))\\
  %  \mathrm{s.t.}\,\,& \hat {\vect s}_{k+1}=\vect f_{\vect\theta}(\hat {\vect s}_k,\hat{ \vect a}_k),\,\,\hat {\vect s}_0=\vect s\\
  %  & \hat{ \vect a}_k\in \mathcal{U}
%\end{align}
%\end{subequations}
%MPC-scheme \eqref{eq:MPC} is a very simple version of a function approximator and we will extend it in the following to enforce constraints satisfactions and closed-loop stability for the model. Indeed ENMPC-scheme \eqref{eq:MPC} with policy \eqref{eq:pol} is obtained by using model \eqref{eq:det:model} in OCP \eqref{eq:value2} with parameterized stage cost and terminal cost.
%\begin{Corollary}\label{Cor1} If there exists a $\vect\theta^\star$ such that $\hat L(\vect s,\vect a)=L_{\vect\theta^\star}(\vect s,\vect a)$ and $\hat T(\vect s)=T_{\vect\theta^\star}(\vect s)$, $\forall \vect s\in\mathcal{S},\vect a\in\mathcal{U}$, then the optimal policy of MPC \eqref{eq:MPC} is equal to the optimal policy of the MDP with either discounted and undiscounted optimality criteria. 
%\end{Corollary}
Due to the mismatch between the model and the real system, hard constraints in the  MPC scheme could become infeasible. This is a well-known issue in the MPC community and one simple solution consists in formulating the state constraints as soft constraints~\cite{kerrigan2000soft}. We therefore formulate the MPC finite-horizon OCP as:
\begin{subequations}\label{eq:MPC2}
\begin{align}
    \hat{V}^{\vect\theta}_N(\vect s)=\min_{\hat{\vect a},\hat{\vect s},\vect \sigma}\,\,& -\lambda_{\vect\theta}(\hat {\vect s}_0)+T_{\vect\theta}(\hat {\vect s}_N)+\vect \mu_{\mathrm{f}}^\top \vect \sigma_{N}\nonumber\\&\qquad\qquad+\sum_{k=0}^{N-1}L_{\vect\theta} (\hat{ \vect s}_k,\hat {\vect a}_k)+\vect \mu^\top \vect \sigma_{k}\label{eq:costMPC}\\
    \mathrm{s.t.}\,\,& \hat {\vect s}_{k+1}=\vect f_{\vect\theta}(\hat {\vect s}_k,\hat {\vect a}_k),\,\,\hat {\vect s}_0=\vect s, \label{eq:con1}\\
    & \hat {\vect a}_k\in \mathcal{U}, \,\,\, 0\leq \vect \sigma_{k}, \,\,\, 0\leq \vect\sigma_{N}, \\
    &\vect h_{\vect\theta}(\hat {\vect s}_k,\hat {\vect a}_k)\leq \vect\sigma^\star_k, \,\,\, \vect h^{\mathrm{f}}_{\vect\theta}(\hat {\vect s}_N)\leq \vect\sigma^\star_N, \label{eq:con3}
\end{align}
\end{subequations}
where $\hat{V}^{\vect\theta}_N$ is the MPC-based parameterized value function, $\vect h_{\vect\theta}(\vect s,\vect a)$ is a mixed input-state constraint, $\vect h^{\mathrm{f}}_{\vect\theta}(\vect s)$ is the terminal constraint, $\vect\sigma_k$ and $\vect\sigma_N$ are slack variables guaranteeing the feasibility of the MPC scheme and $\vect \mu$ and $\vect \mu_{\mathrm{f}}$ are constant vectors that ought to be selected sufficiently large~\cite{kerrigan2000soft}. Note that these constants allow the MPC scheme to find a feasible solution, but penalize constraint violations enough to guarantee that a feasible solution is found whenever possible. While alternative feasibility-enforcing strategies, e.g., robust MPC, do exist, an exhaustive discussion on the topic is beyond the scope of this paper. Function $\lambda_{\vect\theta}$ parameterizes the so-called storage function, which has been added to the cost in order to enable the MPC scheme to tackle the case of so-called economic problems. Such situations arise when the MDP stage cost is not positive definite, while the MPC stage cost is forced to be positive definite in order to obtain a stabilizing feedback policy. 
Note that since the term $-\lambda_{\vect\theta}(\hat {\vect s}_0)$ only depends on the current state, it does not modify the optimal policy. For more details, we refer the interested readers to \cite{Arash2021verification,gros2019data}.

While Theorem~\ref{theorem1} states that one can find suitable stage and terminal costs for any given model, adjusting the model parameters is not essential from the theoretical perspective. However, in practice, the stage and the terminal cost parameterization may not capture $\hat L$ and $\hat T$ exactly. Since $\hat L$ and $\hat T$ are (implicitly) functions of the model, using a parameterized model $\vect f_{\vect\theta}$ introduces extra degrees of freedom to bring $\hat L$ and $\hat T$ closer to the functions that can be represented by $L_{\vect\theta}$ and $T_{\vect\theta}$. In turn, this can yield a better approximation of the optimal policy and value function. The MPC parameterized policy can be obtained from \eqref{eq:MPC2} as follows:
\begin{align}\label{eq:pol}
    \hat{\vect\pi}^{\vect\theta}_N(\vect s)=\hat{\vect a}^\star_{0}(\vect\theta,\vect s),
\end{align}
where $\hat{\vect a}^\star_{0}$ is the solution of \eqref{eq:MPC2}, corresponding to the first input  $\hat{\vect a}_0$. Moreover, the parameterized action-value function based on MPC scheme \eqref{eq:MPC2} can be formulated as follows:
\begin{align}\label{eq:QMPC}
    \hat Q^{\vect \theta}_N(\vect s,\vect a):= \min_{\hat{\vect a},\hat{\vect s},\vect \sigma}\,\, \eqref{eq:costMPC}\,\,,\qquad
    \mathrm{s.t.}\,\, \eqref{eq:con1}-\eqref{eq:con3}\,,\,\, \hat {\vect a}_0= \vect a\,.
\end{align}
Then one obtains the following identities:
\begin{align}
     \hat{V}^{\vect \theta}_N(\vect s)=\min_{\vect a} \hat Q^{\vect \theta}_N(\vect s,\vect a),\,\,
     \hat{\vect\pi}^{\vect\theta}_N(\vect s)\in\mathrm{arg}\min_{\vect a} \hat Q^{\vect \theta}_N(\vect s,\vect a)\,.
\end{align}

We can use RL techniques, such as Q-learning and policy gradient method to tune the parameters $\vect\theta$ of parameterized MPC scheme \eqref{eq:MPC2} and approach the \textit{optimal} parameter $\vect\theta^\star$. For instance, at each learning step, Q-learning based on Temporal difference (TD) method uses the following update rule for $\theta$:
\begin{subequations}\label{eq:Qlearning}
	\begin{gather}
	\delta_{k}:=\ell(\vect s_{k},\vect a_{k})+\gamma \hat V_N^{\vect \theta}(\vect s_{k+1})-\hat Q_N^{\vect \theta}(\vect s_{k},\vect a_{k})\\
	\vect \theta\leftarrow \vect \theta+\zeta\delta_{k}\nabla_{\vect \theta}\hat Q_N^{\vect \theta}(\vect s_{k},\vect a_{k}) \label{Qlearning}
	\end{gather}
\end{subequations}
in order to capture the optimal value function $\hat Q_N^{\vect\theta^\star}\approx Q^\star$ for the optimal parameters $\vect\theta^\star$, where the scalar $\zeta>0$ is the learning step-size, $\delta_k$ is labelled the TD error. The use of RL for the tuning the MPC scheme can be found e.g., in \cite{gros2019data,kordabad2021reinforcement}.

%The next section provides an analytical case study to illustrate the theoretical developments of this paper.

\section{Analytical Case Study}\label{sec:analyt}
We consider a Linear Quadratic Regulator (LQR)  example in order to obtain the corresponding optimal value functions analytically and verify Theorem~\ref{theroem2}. The real system state transition and stage cost are given as follows:
\begin{align}
    \vect s^+=A\vect s+B\vect a+\vect e,\quad \ell(\vect s,\vect a)=\begin{bmatrix}
\vect s\\\vect a 
\end{bmatrix}^\top\begin{bmatrix}
T & N\\N^\top & R 
\end{bmatrix} \begin{bmatrix}
\vect s\\\vect a 
\end{bmatrix},
\end{align}
where $\vect e\sim \mathcal{N}(0,\Sigma)$ with the discount factor $\gamma$. One can verify the following optimal value functions:
\begin{align}\label{eq:value:lqr}
    V^\star(\vect s)&=\vect s^\top S \vect s+\hat v_\infty,\qquad
    \\\nonumber Q^\star(\vect s,\vect a)&=\hat v_\infty+\begin{bmatrix}
\vect s\\\vect a 
\end{bmatrix}^\top\begin{bmatrix}
T+\gamma A^\top S A & N+\gamma A^\top S B\\N^\top+\gamma B^\top S A & R+\gamma B^\top S B
\end{bmatrix} \begin{bmatrix}
\vect s\\\vect a 
\end{bmatrix},
\end{align}
where $\hat v_\infty=\frac{\gamma}{1-\gamma}\mathrm{Tr}(S\Sigma)$ and $S$ is obtained from the following Riccati equations:
\begin{subequations}
\begin{align}
 T+\gamma A^\top S A&=S+(N+\gamma A^\top SB)\left(K_\gamma^\star\right)^\top,\label{eq:ric11}\\
 (R+\gamma B^\top SB)K_\gamma^\star&=N^\top +\gamma B^\top SA.\label{eq:ric12}
\end{align}
\end{subequations}
Then $\vect\pi^\star(\vect s)=-K_\gamma^\star\vect s$ and $\bar{\vect\pi}^\star(\vect s)=\tilde{\vect\pi}^\star(\vect s)=-K_1^\star\vect s$, where $K_1^\star=\lim_{\gamma\rightarrow 1}K_\gamma^\star$. We then consider a linear deterministic model:
\begin{align}\label{eq:model:LQR}
    {\vect s}^+=&\hat A\vect s+\hat B\vect a,
\end{align}
and an undiscounted OCP with the following stage cost, defined according to Equation~\eqref{eq:lhat0} as:
\begin{align}\label{eq:lht:lqr0}
  &\hat L(\vect s,\vect a)= Q^\star(\vect s,\vect a)-  V^\star( \hat{\vect s}^+)\\&\qquad\quad\stackrel{\eqref{eq:value:lqr}}=\begin{bmatrix}
\vect s\\\vect a 
\end{bmatrix}^\top\begin{bmatrix}
T+\gamma A^\top S A & N+\gamma A^\top S B\\N^\top+\gamma B^\top S A & R+\gamma B^\top S B
\end{bmatrix} \begin{bmatrix}
\vect s\\\vect a 
\end{bmatrix}\nonumber\\&-(\hat A\vect s+\hat B\vect a)^\top S (\hat A\vect s+\hat B\vect a)\nonumber
  := \begin{bmatrix}
\vect s\\\vect a 
\end{bmatrix}^\top\begin{bmatrix}
\hat T & \hat N\\\hat N^\top & \hat R 
\end{bmatrix} \begin{bmatrix}
\vect s\\\vect a 
\end{bmatrix}.
\end{align}
The Riccati equations for the undiscounted problem with the model~\eqref{eq:model:LQR} read as:
\begin{subequations}\label{eq:ric:model}
\begin{align}
 \hat T+\hat  A^\top \hat S \hat A&=\hat S+(\hat N+ \hat A^\top \hat S\hat B)\left(\hat K^\star\right)^\top,\\
 (\hat R+\hat  B^\top \hat S\hat B)\hat K^\star&=\hat N^\top +\hat  B^\top \hat S\hat A.
\end{align}
\end{subequations}
with the optimal policy $\hat {\vect\pi}^\star_\infty(\vect s)=-\hat K^\star\vect s$ and the optimal value function $\hat V_\infty^\star(\vect s)=\vect s^\top \hat S \vect s$. From \eqref{eq:lht:lqr0}, we have:
\begin{subequations}
\begin{align}
T+\gamma A^\top S A-\hat A ^\top S   \hat A&=\hat T, \\ 
N+\gamma A^\top S B-\hat A ^\top S   \hat B&=\hat N,\\ 
R+\gamma B^\top S B-\hat B ^\top S   \hat B&=\hat R.
\end{align}
\end{subequations}
Equivalently, this entails that $\hat T$, $\hat N$ and $\hat{R}$ must satisfy
\begin{subequations}
\begin{align}
\hat T+\hat A ^\top S   \hat A&=T+\gamma A^\top S A,\label{eq:hat1} \\ 
\hat N+\hat A ^\top S   \hat B&=N+\gamma A^\top S B,\label{eq:hat2} \\ 
\hat R+\hat B ^\top S   \hat B&=R+\gamma B^\top S B.\label{eq:hat3}
\end{align}
\end{subequations}
Then:
\begin{align}\label{eq:ric1}
    &\hat T+\hat A ^\top S   \hat A\stackrel{\eqref{eq:hat1}}=T+\gamma A^\top S A\stackrel{\eqref{eq:ric11}}=S+\\&\,\,\,S(N+\gamma A^\top SB)\left(K_\gamma^\star\right)^\top\stackrel{\eqref{eq:hat2}}=S+(\hat N+\hat A ^\top S   \hat B)\left(K_\gamma^\star\right)^\top,\nonumber
\end{align}
and
\begin{align}\label{eq:ric2}
    (\hat R+\hat B^\top S\hat B)K_\gamma^\star&\stackrel{\eqref{eq:hat3}}=(R+\gamma B^\top SB)K_\gamma^\star\\&\stackrel{\eqref{eq:ric12}}=N^\top +\gamma B^\top SA\stackrel{\eqref{eq:hat2}}=\hat N+\hat A ^\top S   \hat B.\nonumber
\end{align}
Equations \eqref{eq:ric1} and \eqref{eq:ric2} show that $\hat S=S$ and $\hat K^\star=K_\gamma^\star$ satisfy the undiscounted Riccati equations \eqref{eq:ric:model}. Then it reads that $\vect\pi^\star(\vect s)=\hat{\vect\pi}_\infty^\star(\vect s)$ and $V^\star(\vect s)=\hat V_\infty^\star(\vect s)+\hat v_\infty$. 
\subsection{Satisfying the assumptions}\label{sec:assum}
\begin{figure}[t]
    \centering
   \includegraphics[width=0.3\textwidth]{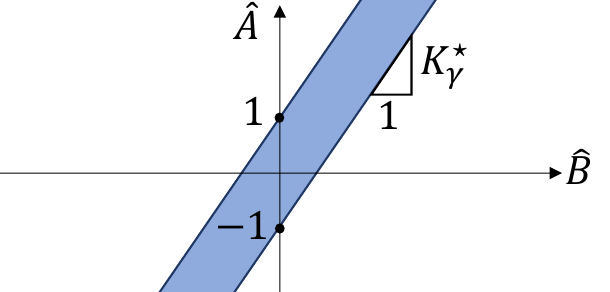}
    \caption{The blue area shows all $\hat{A}$ and $\hat{B}$ in the linear model such that the resulting trajectory and optimal value function remain bounded for the given optimal policy $\vect\pi^\star(\vect s)=-K_\gamma^\star\vect s$.}
    \label{fig:assum}
\end{figure}

Regarding Assumption~\ref{assume:S}, the value function will remain bounded in the finite horizon prediction for every bounded initial condition $\vect s_0$ and every linear model in form \eqref{eq:model:LQR} for a given control policy $\vect\pi^\star(\vect s)=-K_\gamma^\star\vect s$ or $\bar{\vect\pi}^\star(\vect s)=\tilde{\vect\pi}^\star(\vect s)=-K_1^\star\vect s$. For Assumption~\ref{assume2}, the linear model matrices $\hat{A}$ and $\hat{B}$  must be chosen such that
$\rho(\hat{A}-\hat{B}K_\gamma^\star)\leq 1$ in order to guarantee boundedness of the optimal value function \eqref{eq:value:lqr}. For instance, for a scalar dynamics, the locus of $\hat{A}$ and $\hat{B}$ is shown in Figure~\ref{fig:assum}.
Inspired by this example, we ought to point out here that for linear systems Assumption~\ref{assume:S} is automatically obtained if the model is stabilized by the optimal policy, though the converse might not be true (e.g., if the cost is $0$). Note that, the systems without constraint satisfying Assumption~\ref{assume:S} is fairly straightforward while in the presence of the system constraints, the model also must not violate those constraints. To satisfy Assumption~\ref{assume2}, a model must be adopted whose trajectory does not diverge under the optimal policy of the real system and satisfy the system constraint. It is clear that the closer the model is to the real system the more likely it is to satisfy this assumption. This model can be obtained based on offline system identification. In \cite{zanon2020safe}, the authors proposed to use robust MPC in order to ensure constraint satisfaction. A deeper discussion of these assumptions can be found in \cite{gros2019data} and \cite{zanon2022stability}.
\section{Numerical Examples}\label{sec:sim}
\subsection{Non-quadratic stage cost}
\begin{figure}[t]
\centering
\includegraphics[width=0.48\textwidth]{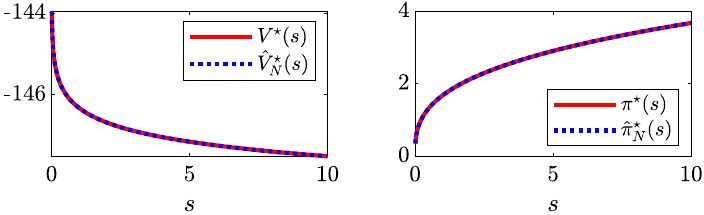}
\caption{(Left:) Optimal value functions (Right:) and optimal policy resulting from the discounted real system and undiscounted MPC scheme with the wrong model.}
\label{fig:F1}
\end{figure}

In this example, we provide a benchmark optimal investment problem with a non-quadratic stage cost. Consider the following dynamics and stage cost~\cite{santos1998analysis}:
\begin{align}\label{eq:non:pol}
    s_{k+1}=a_k \,, \qquad \ell(s,a)=-\ln(As^{\alpha}-a),\, 
\end{align}
where $A$ and $0<\alpha<1$ are given constants. It is known that for the discount factor $\gamma$, the optimal value and policy functions are $V^\star(s)=B+C\ln(s)$ and $\pi^\star(s)=\gamma\alpha A s^{\alpha}$, where~\cite{grune2016discounted}:
\begin{align}
B=\frac{\ln((1-\alpha\gamma)A)+\frac{\gamma\alpha}{1-\gamma\alpha}\ln(\alpha\gamma A)}{\gamma-1},\quad
C=\frac{\alpha }{\alpha\gamma-1}\,.
\end{align}
We then consider a model of the dynamics with $\hat s_{k+1}=\mu \hat a_k$ and, based on this model, we construct a finite-horizon undiscounted MPC  with the costs according Equation~\eqref{eq:TL} in Theorem~\ref{theorem1} and $N=10$. In this example we have considered $A=5$, $\alpha=0.34$, $\mu=0.8$ and $\gamma=0.9$. Figure~\ref{fig:F1} compares the optimal value and policy functions from the discounted real system \eqref{eq:non:pol} and from the MPC scheme with a wrong model. As predicted by Theorem~\ref{theorem1}, one can see that they match perfectly. Note that the results are valid for every discount factor $0<\gamma<1$, every horizon length and for other values of the constants $A$, $\alpha$, and $\mu$.

%\begin{figure}[t]
%\centering
%\includegraphics[width=0.4\textwidth]{F2.pdf}
%\caption{Optimal policy functions resulting from the discounted real system and undiscounted MPC scheme with the wrong model.}
%\label{fig:F2}
%\end{figure}

\subsection{Inverted pendulum with process noise}
\begin{figure}[t]
\centering
\includegraphics[width=0.48\textwidth]{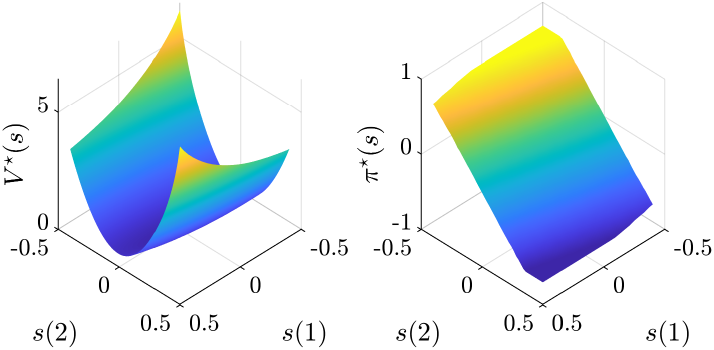}
\caption{Optimal Value (left) and policy (right) functions resulting from ADP.}
\label{fig:V}
\end{figure}

We consider the following discrete-time stochastic dynamics, representing an inverted pendulum with a random support excitation:
\begin{align}\label{eq:dyn:ex}
    \vect s_{k+1}=\vect s_{k}+\begin{bmatrix}
 s_k(2)\\(\frac{g}{l}+\xi)\sin(s_k(1))
\end{bmatrix}\delta t+\begin{bmatrix}
 0\\\frac{\delta t}{ml^2}
\end{bmatrix} \vect a_k
\end{align}
where $g=9.81$, $l=0.3$, $m=0.5$ and $\delta t=0.1$ are constants representing the gravity, mass, length and the sampling time of the discrete dynamics. Disturbance $\xi\sim\mathcal{U}[-0.5,0.5]$ has a uniform distribution and $\vect s_k:=[s_k(1),\, s_k(2)]^\top$ is the system state and $\vect a_k$ is the system input. We consider $\ell(\vect s,\vect a)=\vect s^\top \vect s+\vect a^2$ as a stage cost with the discount factor $\gamma=0.95$. We first aim to find an approximate solution for the optimal policy and the optimal value functions using Dynamic Programming (DP). We consider the state  constraints $-1\leq s_k(1)\leq 1$, $-1\leq s_k(2)\leq 1$ and the input constraint $-0.8\leq a_k\leq 0.8$. Figure~\ref{fig:V} shows the optimal value function and the optimal policy function resulting from DP for the discounted infinite-horizon MDP.

\begin{figure}[t]
\centering
\includegraphics[width=0.48\textwidth]{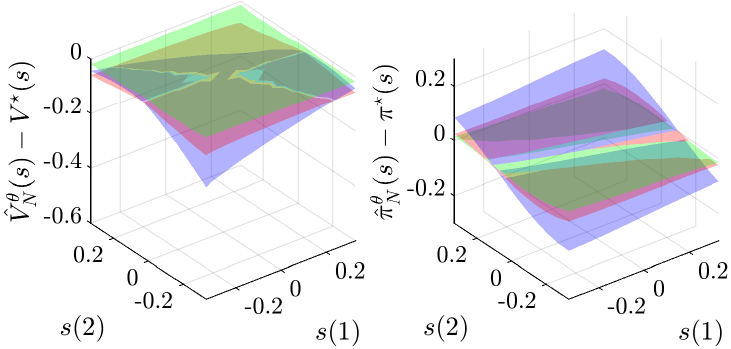}
\caption{The difference between the MPC based parameterized value  (left)\textbackslash policy (right) and their optimal solutions for the beginning of the learning (blue) and after $500$ learning steps (red) and the exact cost modification from theorem~\ref{theorem1} (green).}
\label{fig:Vd}
\end{figure}

%\begin{figure}[t]
%\centering
%\includegraphics[width=0.4\textwidth]{F4.pdf}
%\caption{Optimal Policy function resulting from DP.}
%\label{fig:PI}
%\end{figure}
We build an undiscounted finite-horizon OCP with a wrong model in order to capture the optimal value and the optimal policy functions of the discounted infinite horizon MDP. To do this, we consider an MPC scheme with a deterministic linearized form of the dynamics as a model of the real system as follows:
\begin{align}\label{eq:Model:ex}
    \hat{\vect s}_{k+1}=\vect f_{\vect\theta}(\hat{\vect s}_{k},\hat{\vect a}_k)=\hat{\vect s}_{k}+\begin{bmatrix}
 \hat{s}_k(2)\\\frac{g}{\theta_l}\hat{s}_k(1)
\end{bmatrix}\delta t+\begin{bmatrix}
 0\\\frac{\delta t}{m\theta_l^2}
\end{bmatrix} \hat{\vect a}_k
\end{align}
where $\hat{\vect s}_{k}:=[\hat s_k(1),\,\hat s_k(2)]^\top$ and $\hat{\vect a}_k$ are the model state and input. Moreover, we consider an uncertain $l$ with a adjustable parameter $\theta_l$, with an initial value $0.25$.
We consider the parameterized MPC scheme with the horizon length $N=10$ and the following parameterized quadratic stage and terminal cost:
\begin{align}\label{eq:quad:cost:ex}
    T_{\vect\theta}(\vect s)=\vect s^\top G\vect s,\qquad
    L_{\vect\theta}(\vect s,\vect a)=\begin{bmatrix}
\vect s\\a 
\end{bmatrix}^\top H
\begin{bmatrix}
\vect s\\a 
\end{bmatrix}
\end{align}
where $G$ and $H$ are parametric positive definite matrices. Then the parameters vector $\vect\theta$ gathers all the adjustable parameters as $\vect\theta=\{\theta_l,\,G,\,H\}$. We use the  Q-learning method in order to update the parameters $\vect\theta$ to achieve the optimal solutions of the real system and improve the closed-loop performance. Figure~\ref{fig:Vd} shows the difference between the MPC value $\hat{V}^{\vect\theta}_N$ and policy $\hat{\vect\pi}^{\vect\theta}_N$ functions with their optimal solutions computed by DP. The blue and red surfaces represent this difference at the beginning of the learning and after $500$ learning steps, respectively. As it can be seen, the results are getting closer to zero as the learning proceeds. 
Note that the stage and terminal costs yielding a perfect match of $V^\star$ and $\vect\pi^\star$, as per Theorem~\ref{theorem1}, do not have a quadratic form, hence the selected MPC formulation cannot capture them exactly. The green surfaces in Figure~\ref{fig:Vd} have been obtained by computing these stage and terminal costs numerically and shows the corresponding $\hat V^\star_N- V^\star$ and $\hat {\vect\pi}^\star_N-{\vect\pi}^\star$. As expected the difference is zero, modulo tiny numerical inaccuracies. 

\begin{figure}[t]
\centering
\includegraphics[width=0.4\textwidth]{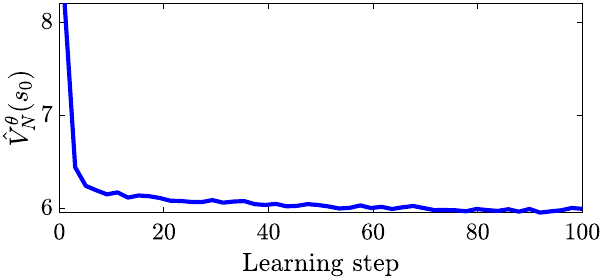}
\caption{The MPC-based value function $\hat V^{\vect\theta}_N(\vect s_0)$ during the learning.}
\label{fig:per}
\end{figure}

%\begin{figure}[t]
%\centering
%\includegraphics[width=0.4\textwidth]{F6.pdf}
%\caption{The difference between the MPC based parameterized policy function and optimal policy function for the beginning of the learning (blue) and after $500$ learning steps (red) and $\hat {\vect\pi}^\star_N- {\vect\pi}^\star$ (green).}
%\label{fig:PId}
%\end{figure}
Finally, Figure~\ref{fig:per} illustrates the closed-loop performance of the system under the MPC policy $\hat{\vect\pi}^{\vect\theta}_N$. 
As the closed loop cost decreases, this demonstrates that RL can be effective in tuning the MPC parameters so as to achieve the best closed-loop performance.

\subsection{Learning based MPC: Tracking stage cost}

\begin{figure}[t]
	\centering
	{\def\svgwidth{0.39\textwidth}
			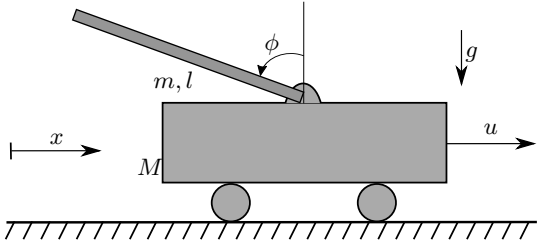
			}
	\caption{The cart-pendulum system. We use $M=0.5\mathrm{kg}$, $m=0.2\mathrm{kg}$, $l=0.3\mathrm{m}$ and $g=9.8\mathrm{m/s^2}$ for the simulation.}
	\label{figdyn}
\end{figure}

\begin{figure}[t]
\centering
\includegraphics[width=0.4\textwidth]{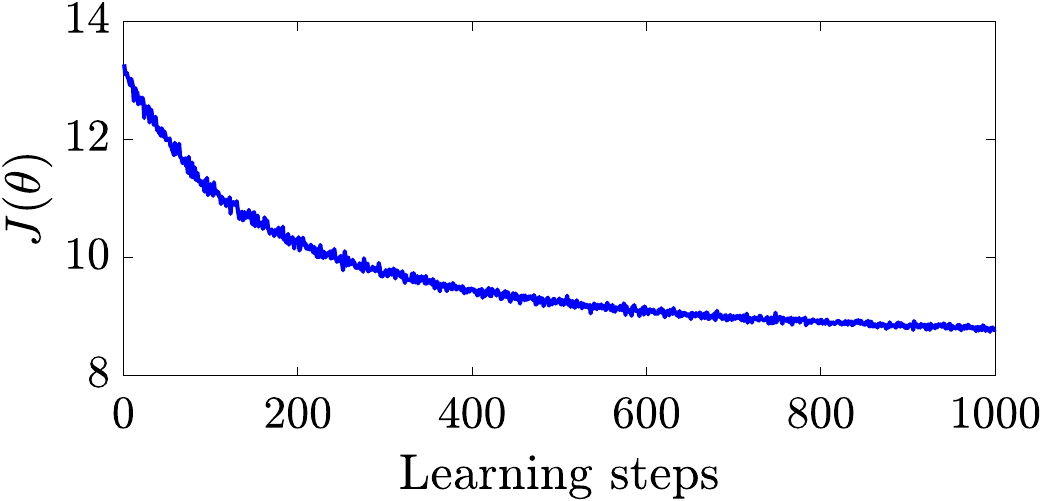}
\caption{The closed-loop performance of the MPC scheme over RL-steps.}
\label{fig:1}
\end{figure}

In this section, we consider the cart-pendulum balancing problem shown in Figure~\ref{figdyn} in order to illustrate the proposed method in a constrained tracking problem. The dynamics are given by:
\begin{subequations}\label{eq:cp}
\begin{align}
  (M+m)\ddot{x}+\frac{1}{2}ml\ddot{\phi}\cos\phi &=\frac{1}{2}ml\dot{\phi}^2\sin\phi+u,\\
  \frac{1}{3}ml^2\ddot{\phi}+\frac{1}{2}ml\ddot{x}\cos\phi &=-\frac{1}{2}mgl\sin\phi,
\end{align}
\end{subequations}
where $M$ and $m$ are the cart mass and pendulum mass, respectively, $l$ is the pendulum length and $\phi$ is its angle from the vertical axis. Force $u$ is the control input, $x$ is the cart displacement and $g$ is gravity. 
We use the Runge-Kutta $4^{\mathrm{th}}$-order method to discretize \eqref{eq:cp} with a sampling time $\mathrm{d}t=0.1\mathrm{s}$ and cast it as $\vect s^+=\vect f(\vect s,\vect a)+\vect\xi$, where $\vect s=[ x,\dot x, \phi,\dot \phi]^\top$ is the state, $\vect a=u$ is the input, $\vect\xi$ is a Gaussian noise and $\vect f$ is a nonlinear function representing \eqref{eq:cp} in discrete time. We consider the state constraint $x\geq 0$, discount factor $\gamma=0.95$ and the following MDP stage cost to stabilize the system at the origin while penalizing the system constraint:
\begin{align}
        \ell(\vect s,\vect a)=& \begin{bmatrix}
\vect s\\\vect a 
\end{bmatrix}^\top\begin{bmatrix}
I_4 & 0\\0 & 0.01 
\end{bmatrix} \begin{bmatrix}
\vect s\\\vect a 
\end{bmatrix}+ \lambda \mathrm{max}(-x,0),
\end{align}
where $\lambda$ is a large constant value introduced to model the state constraint as a soft constraint. In the MPC scheme, we use the linear model ${\vect s}^+=\hat A{\vect s}+\hat B{\vect a}$ obtained by linearizing $\vect f$ at the origin. We provide a parametrized quadratic stage and terminal cost and select prediction horizon $N=20$. We use the deterministic policy gradient method to minimize the performance function $J(\vect\theta):=\mathbb{E}_{\vect s_0}[\hat V_N^{\vect{\theta}}(\vect s_0)]$, and we run a simulation for $1000$ learning steps of the policy gradient method. Figure~\ref{fig:1} shows the value function over the learning steps for a fixed initial state. This illustrates that RL successfully manages to reduce $J$ throughout the iterates, therefore tuning MPC as desired.

\begin{figure}[t]
\centering
\includegraphics[width=0.4\textwidth]{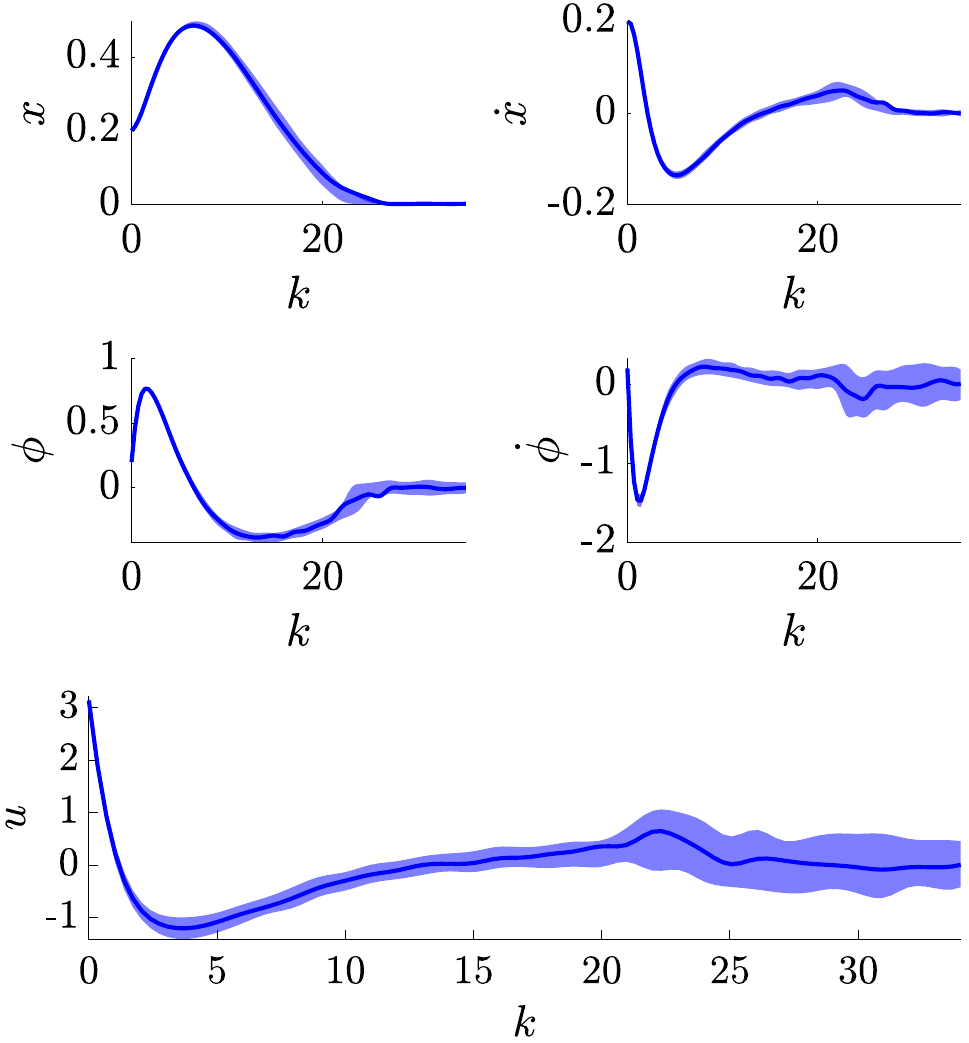}
\caption{States and input trajectories of the real system for the last learning step.}
\label{fig:2}
\end{figure}

\begin{figure}[t]
\centering
\includegraphics[width=0.35\textwidth]{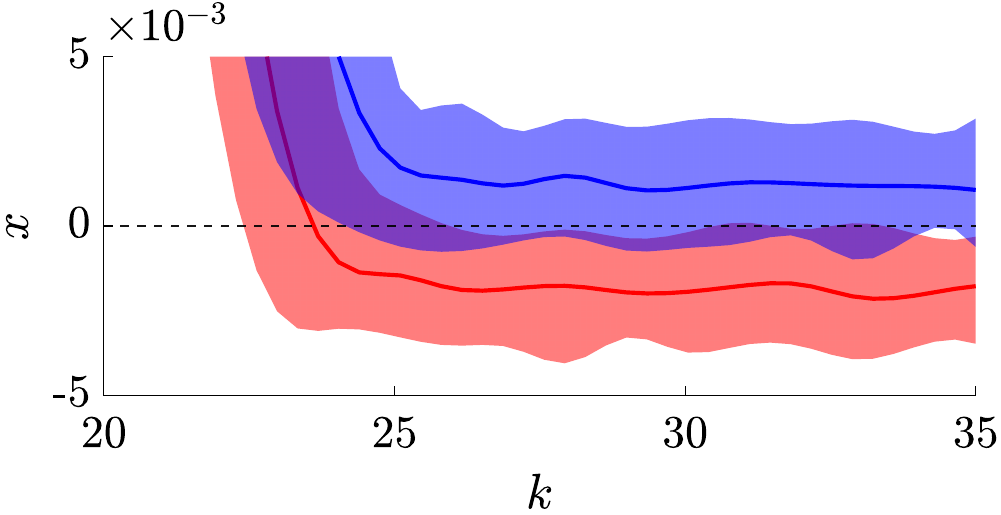}
\caption{Violation of the state constraint $x\geq 0$ in the first step (red) and the last step (blue).}
\label{fig:3}
\end{figure}

Figure~\ref{fig:2} shows the states and input trajectories of the real system corresponding to the $1000^{\mathrm{th}}$ learning step of the policy gradient method. The MPC scheme with the positive definite stage cost and other stability conditions in the terminal cost, terminal constraint is able to deliver the stabilizing policy for the closed-loop system for the small enough model error~\cite{MPCbook}. Note that the terminal cost and constraint conditions can be relaxed for the large enough MPC horizon~\cite{jadbabaie2005stability}. Figure~\ref{fig:3} compares the state constraint violation for $x\geq 0$ in the first and the last ($1000^{\mathrm{th}}$) learning step. As one can see, RL reduces the state constraint violation. Note that, we have used a common MPC formulation as~\eqref{eq:MPC2} in this example. However, one can use robust MPC to avoid constraint violation as shown in~\cite{zanon2020safe}.

\subsection{Learning based MPC: Economic stage cost}

\begin{figure}[t]
\centering
\includegraphics[width=0.48\textwidth]{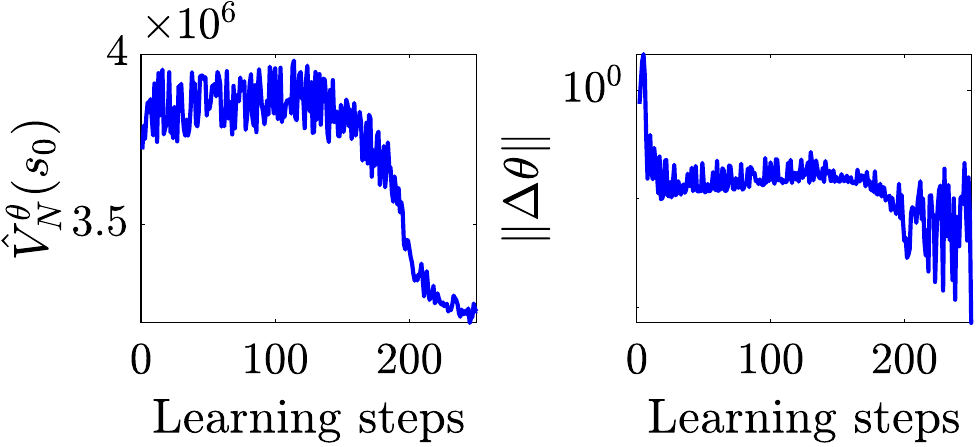}
\caption{(Left:) The MPC-based value function $\hat V^{\vect\theta}_N(\vect s_0)$ during the learning .(Right:) Convergence of the norm of the parameters during the Q-learning steps.}
\label{fig:4}
\end{figure}

In this example, we investigate an economic cost in the real system with bias optimality criterion. We use a parameterized MPC scheme with a parameterized storage function as a function approximator in the Q-learning algorithm. Continuously Stirred Tank Reactor (CSTR) is a common ideal reactor in chemical engineering, usually used for liquid-phase or multiphase reactions with fairly high reaction rates. The CSTR nonlinear dynamics can be written as follows (see ~\cite{Economic2016CSTR}):
\begin{align}\label{eq:dyn:CSTR}
\dot{C}_A&=\frac{F}{V_R}(C_{A0}-C_A)-k_0 e^{-{E}/{RT}}C_A^2 \\
\dot T&=\frac{F}{V_R}(T_0-T)-\frac{\Delta H k_0}{\rho_R C_p}e^{-{E}/{RT}}C_A^2+\frac{q}{\rho_R C_p V_R}, \nonumber
\end{align}
where $T$ denotes the temperature of the reactor contents, $C_A$ is the concentration of $A$ in the reactor, $F$ is the flow rate, and $q$ is the heat rate. The remaining notation definitions and process parameter values are given in e.g., \cite{kordabad2022q}. 
%\begin{table}[t]
%\caption{\label{tab:table1} Parameter definitions and values of CSTR.}
%\centering
%\begin{tabular}{ccc}
%\hline
%Symbol    & Description & Value\\
%\hline
%$C_{A0}$      & Feed concentration of $A$    & 3.5$kmol/m^3$ \\
%$T_0$       & Feedstock temperature     & 300$K$      \\
%$V_R$       & Reactor fluid volume     & 1.0$m^3$      \\
%$E$ & Activation energy      & 5.0e4$kJ/kmol$       \\
%$k_0$ & Pre-exponential rate factor      & 8.46e6$m^3/kmolh$       \\
%$\Delta H$ & Reaction enthalpy change      & -1.16e4$kJ/kmol$       \\
%$C_p$ & Heat capacity      & 0.231$kJ/kgK$       \\
%$\rho_R$ & Density      & 1000$kg/m^3$       \\
%$R$ & Gas constant      & 8.314$kJ/kmolK$   \\
%\hline
%\end{tabular}
%\end{table}
Then $\vect s=[C_A\,,\, T]^\top$ and $\vect a=[F \,,\, q]^\top$ are the  state and input of the system, respectively. The input $\vect a$ must satisfy the following inequality: 
\begin{align}
  [0\,,\,-2\mathrm{e}5]^\top \leq  \vect a \leq [10\,,\,2\mathrm{e}5]^\top
\end{align}
An economic stage cost is defined as follows:
\begin{align}
    \ell(\vect s,\vect a)= -\eta \underbrace{F (C_{A0}-C_A)}_{:=r}+\beta q
\end{align}
where $\eta$ and $\beta$ are positive constants, and $r$ is the production rate. This cost maximizes the production rate and minimizes the energy consumption of the production (the second term). We consider $\eta= 1.7\mathrm{e}4$ and $\beta= 1$ for the simulation. Sampling time $0.02\mathrm{h}$ is used to discretize the system \eqref{eq:dyn:CSTR}. 
%The gain optimal value function can be obtained as follows:
%\begin{align}
 %   \bar V^\star=\ell(\vect s_e,\vect a_e) 
%\end{align}
%\textcolor{red}{to do: double check! Is average cost equal to stage cost at the optimal steady state point?}
%where $(\vect s_e,\vect a_e)$ is an optimal steady-state pair and is calculated as follows:
%\begin{align}
 %   \vect s_e=[0.7572\,,\, 497.71]^\top \,,\, \vect a_e=[10.00 \,,\, 1.38557\mathrm{e}5]^\top
%\end{align}
We use an MPC scheme with a neural network-based storage function and parameterized stage cost and terminal cost and we denote the adjustable parameters by $\vect\theta$. Then we use Q-learning in order to update the parameters $\vect\theta$. Figure~\ref{fig:4} (left) illustrates the value function $\hat V^{\vect\theta}_N(\vect s_0)$. It can be seen that the parameterized value function is decreasing during the learning. Figure~\ref{fig:4} (right) shows the convergence of the parameters.

%\begin{figure}[t]
%\centering
%\includegraphics[width=0.4\textwidth]{F13.pdf}
%\caption{The MPC-based value function $\hat V^{\vect\theta}_N(\vect s_0)$ during the learning.}
%\label{fig:5}
%\end{figure}
\section{Conclusion}\label{sec:cons}
In this paper, we showed that a finite-horizon OCP can capture the optimal policy and value functions of any MDPs with either discounted or undiscounted cost even if we use an inexact model in the OCP. We showed that an MPC scheme can be interpreted as a particular case of the OCP where we use a deterministic model to avoid computational complexity. In practice, we proposed the use of a parameterized MPC scheme to provide a structured function approximator for the RL techniques. RL algorithms then can be used in order to tune the MPC parameters to achieve the best closed-loop performance. We verified the theorems in an LQR case and investigated some nonlinear examples to illustrate the efficiency of the method numerically.
\typeout{}
%\section*{Acknowledgment}
%This work was funded by the Research Council of Norway (RCN) project ``Safe Reinforcement Learning using MPC" (SARLEM).

\bibliographystyle{IEEEtran}
\bibliography{EMDP}

\end{document}